\documentclass{PoS}
\usepackage{ntheorem}
\newtheorem{ass}{Assumption}
\newtheorem{ass2}{Assumption}
\usepackage{epigraph}
\usepackage{soul} 
\usepackage{graphicx}
\usepackage[numbers,sort&compress]{natbib} 

\usepackage{amsmath}
\usepackage[utf8]{inputenc}
\usepackage{etoolbox}
\makeatletter
\patchcmd{\epigraph}{\@epitext{#1}}{\itshape\@epitext{#1}}{}{}
\makeatother

\newcommand{\lag}{\mathcal{L}}
\newcommand{\op}{\mathcal{O}}
\newcommand{\amp}{\mathcal{A}}

\title{Effective field theories and pseudo-observables in the quest for physics beyond the Standard Model}

\ShortTitle{Effective Field Theories and Pseudo Observables}

\author{\speaker{R. Gomez-Ambrosio}%
         \thanks{Work done in collaboration with M. Ghezzi, G. Passarino and S. Uccirati. Supported by the Executive Research Agency (REA) of the European Union under the Grant Agreement PITN-GA-2012-31674 (HiggsTools). }\\
        Turin Univ. + INFN\\
        E-mail: \email{raquel.gomez@to.infn.it}}

\abstract{
 We discuss briefly the kappa framework, proposed originally as a test for the Higgs couplings of the Standard Model (SM). Further, we discuss a generalization of this idea in terms of effective field theory. We sketch how to add dimension 6 operators to the SM Lagrangian and the renormalization process. Finally we show how to study the amplitudes of the resulting model at next-to-leading order and discuss possible experimental approaches.    
}

\FullConference{18th International Conference From the Planck Scale to the Electroweak Scale \\
		 25-29 May 2015\\
		 Ioannina, Greece }
		
\begin{document}

\section{Introduction}

The aim of these proceedings is to discuss a theory for the study of Standard Model deviations. This theory was presented in \cite{GPNLOHEFT, YR3, GPshort, trott2015, NLOHEFT} as a next-to-leading order (NLO) extension of the $\kappa$-framework \cite{YR3, kappa}. 

It has been 3 years now since the Higgs candidate was observed in ATLAS and CMS experiments \cite{ATLASHiggs, CMSHiggs} and despite of the big success of the first run of LHC, it is hard to imagine how will new discoveries look like: Will we find light supersymmetric particles at the LHC detectors? Will we see a composite-Higgs? Will we see extra-dimensions? How will they look like? 
 
In LHC RUN-II a per-mille sensitivity for Higgs and electroweak observables is to be expected and the theory community has to sum efforts to reach the same precision in its predictions. Not only in the predictions in the on-shell regions, nearby the resonances, but also in the off-shell regions, where the new beyond-standard-model (BSM) scenarios are more likely to be detected. 
\newline{}
\newline{}
\noindent \textbf{Pseudo-Observables Vs Fiducial Observables}

Fiducial observables are those defined in a certain fiducial volume, i.e. the volume where the detector operates at its highest efficiency. This volume is different for each experiment and even though the sets of cuts are always designed to be minimal, sometimes it is not possible to compare fiducial results, namely \textit{total} cross-sections, between experiments. On the contrary, Pseudo-Observables (POs) are well-defined objects that can be reconstructed from different sets of cuts. For example: couplings, decay widths and masses (electroweak precision data,  most of them already measured at LEP \cite{LEP}). Suggestively one can write: $\rm{FO} = \rm{PO} \oplus \rm{SM}_{\rm{remnant}}$, i.e. the POs are the ``non fiducial'' factor that can be extracted from the fiducial observables.   

An effective field theory (EFT) approach provides us with a useful tool to bring closer the SM and the various BSM models. This is because of its dual nature: on the one hand, on its top-down realization it can be used to integrate out heavy degrees of freedom of a particular BSM model (by means of the covariant derivative expansion, see for example \cite{murayama}) and on the other hand, on its bottom-up realization, it can be used to extend the SM and see where does it take us.

In the calculations sketched here, we present an EFT, in terms of a renormalized Lagrangian, consisting of $\lag_{SM}$ plus some dimension 6 terms. We write amplitudes as combinations of dim. 4 and ``deformed'' subamplitudes, and we compute some relevant processes at NLO, following the hierarchy:

\begin{equation}
\mathcal{M} = \mathcal{M}_{SM}^{\rm{LO}} +\underbrace{\mathcal{M}_{SM}^{\rm{NLO}}}_{\substack{\text{all available}\\\text{QCD, EW corrections}}}  + \underbrace{\mathcal{M}_{dim = 6}^{\rm{LO}} + \mathcal{M}_{dim = 6}^{\rm{NLO}}}_{\substack{\kappa-\text{framework }\\\text{and dim 6 EFT}}}  
\end{equation}


\section{Looking for deviations: The $\kappa$-framework}

\epigraph{``The thing that doesn't fit is the most interesting.''}{Richard. P. Feynman}

The kappa framework \cite{kappa, YR3} was proposed by the LHC-HXSWG\footnote{LHC Higgs Cross Section Working Group} shortly after the discovery of the 125 GeV resonance.
The idea of the kappa framework is to introduce some ad-hoc variations of the coupling constants and decay widths of the SM Higgs, and see how these would fit the experimental data. In particular, in the case of no deviation from the SM, these $\kappa$'s would be equal to 1. In that case where $\kappa \approx 1$ we can say that the extension that is being searched for is very close to  $\lag_{\rm{SM}}$. This approach has the advantage that it is almost model independent\footnote{ Up to certain model dependence introduced by the unfolding process: when we reconstruct the observables from raw data we need to include PDFs.}. Moreover, if one writes an extension of the SM in  effective field theory language, it can be found that those kappa parameters are not so obscure as they might seem: they are combinations of the Wilson coefficients of the new theory.

To understand the motivation under the kappa framework we can compare LHC with LEP experiments: there was a previous phase during LEP and until the measurement of $\rm{M_H}$ at LHC, where the task was to fill-in some blanks in the SM (namely, couplings and masses). However, since the Higgs mass has been measured, the Standard Model is complete, in the sense that there is no degree of freedom left to make predictions. At this point, if we want to make predictions to be tested , we have to think of new parameters for the model, i.e we need to define new sets of observables to be measured, and fitted, moving from the \textit{predictive} phase to the \textit{fitting} one. And this is where POs enter the game.   

The study of couplings is very interesting because they can be extracted directly from Green's functions, in particular they are the residues of the poles after removing 1PI terms \cite{isidoriPOs, GPAndre}. For every process we look at 
(i.e. for every number of legs and vertices), we can have a finite and well-defined set of POs to be measured. Besides from couplings and decay widths, other interesting POs to be measured are the S, T (or $\rho$) and U oblique parameters \cite{STU} and the electroweak mixing angle: $s_\theta$,$c_\theta$. This framework starts from some tight assumptions:

\begin{ass}
The signal observed comes from a single resonance, at $M_H^2 = 125.09 \pm 0.24$ GeV \cite{higgsMass} with the properties of the SM Higgs regarding spin, parity and electroweak symmetry breaking (EWSB) mechanism. 
\end{ass}

\begin{ass}[Zero width approximation]
The width of the resonance around the peak can be neglected\footnote{It can be shown that this is a highly non-trivial assumption, see \cite{GPZWA}}. And we can write, for all channels: 
\begin{equation}
\left( \sigma \cdot \rm{BR} \right) ( ii \rightarrow H \rightarrow ff ) = \frac{\sigma_{ii} \cdot \Gamma_{ff}}{\Gamma_H} 
\end{equation}

\end{ass}


In any gauge theory, coupling constants and masses are strictly related: one cannot alter the former without affecting the latter, unless new terms are added to the Lagrangian, which we will discuss in the next section. As a consequence, new terms in the Lagrangian might affect the tensor structure of the couplings. To avoid this, one further assumption is made:

\begin{ass}
Only variations in the absolute values of the couplings are considered, hence not altering the structure predicted by the SM.  
\end{ass}

For example, for the case of the main production and decay channels, $ \, \,  g g \rightarrow H \rightarrow \gamma \gamma \quad$ we would write, 
\begin{equation}
\left( \sigma \cdot \rm{BR} \right) ( gg \rightarrow H \rightarrow \gamma \gamma ) = \sigma_{\rm{SM}} \cdot \rm{BR}_{SM} \cdot \frac{\kappa_g^2 \, \kappa_\gamma^2 }{ \kappa_H^2}  
\end{equation}

where the SM cross section and branching ratios can be found in \cite{HXSWG} and the kappas are defined, heuristically\footnote{Kappas for the other production and decay channels are defined analogously, see \cite{kappa}}:
\begin{equation}
\kappa_{\rm{g}}^2 = \frac{\sigma_{\rm{ggH}}}{\sigma_{\rm{ggH}}^{\rm{SM}}}, \qquad 
\kappa_{\gamma}^2 = \frac{\Gamma_{\gamma \gamma}}{\Gamma_{\gamma \gamma}^{\rm{SM}}}, \qquad \kappa_{\rm{H}}^2 = \frac{\Gamma_{\rm{H}}}{\Gamma_H^{\rm{SM}}} .
\end{equation}
Note that both production and decay processes are not at tree level, but contain a loop, and we treat them as \textit{effective couplings}. Therefore, all possible intermediate particles have to be taken into account when we define them, for example:
\begin{equation}
\kappa_{\rm{g}}^2 = \frac{\kappa_t^2 \sigma_{ggh}^{tt}(m_H) + \kappa_b^2 \sigma_{ggh}^{bb}(m_H) + \kappa_t \kappa_b \sigma_{ggh}^{tb}(m_H)  }{\sigma_{ggh}^{bb}(m_H) + \sigma_{ggh}^{tt}(m_H) + \sigma_{ggh}^{tb}(m_H)    }
\end{equation}

The predictions of the original kappa framework  were tested against the data of LHC-Run I, but no significant deviations were found \cite{ATLAScouplings,Jez:2015wza}. We need more experimental resolution and theoretical accuracy. From the theory side we try to achieve this by generalizing the $\kappa$-framework to NLO order.

\section{Generalized Kappa Framework}

The kappa framework, in its original formulation  is rather limited: it does not respect gauge invariance and unitarity. Also, it does not account for possible changes in the tensor structure of the couplings. We need a theory where spin, CP and couplings are treated coherently. This is addressed in \cite{NLOHEFT}, where an NLO generalization of the kappa framework is developed, always following the spirit that SM deviations must have an SM basis. 

Effective field theories have been historically legitimated in all branches of physics: Landau-Ginzburg theory for superconductivity, $\sigma$-models for ferromagnetism and pion scattering, and even General relativity might be an effective field theory for gravity, as well as Newton theory is.  

For the case we study in particular, the best example is that of Fermi theory for $\beta$-decay: an effective field theory for the electroweak interactions. 
This theory was predictive at a certain energy scale, but did not account for some underlying processes (and particles) that exist at higher energies.

In fact, EFT as a parametrization of BSM physics was already proposed in 1986 by Buchm\"uller and Wyler \cite{BuchmullerWyler}, developed further in \cite{warsaw} and first suggested as natural extension to the Kappa framework in \cite{YR3}.
Additionally this approach seems very appropriate to fit the needs of the theory community: since the same effective Lagrangian can arise from different BSM models (as their low-energy limit), Standard-Model-EFT can be seen as the middle point where BSM theories and the SM meet.

\subsection{Bottom-Up Effective Field Theory}

The advantage of bottom-up EFT with respect to the top-down approach is that it is almost model independent, up to some assumptions:

\begin{ass2}
There is one Higgs doublet, in a linear representation. This is a flexible assumption, see some examples of nonlinear realizations in \cite{GPNLOHEFT, nonlinearEFT}.
\end{ass2}

\begin{ass2}
The EFT does not add new light degrees of freedom, and it does not account for light BSM particles in loops.
\end{ass2}

\begin{ass2}
The heavy degrees of freedom decouple. They do not mix with the Higgs doublet and the theory stays renormalizable after removing them.
\end{ass2}

\begin{ass2}
The UV completion is weakly coupled and renormalizable.
\end{ass2}

Here, we start from the SM and extend it by adding higher dimensional operators. Note that this is the counter-intuitive way of doing EFT: usually one starts from known results and tries to build an oversimplified (effective) theory to reproduce them, in this case, instead, the SM is the effective theory, and we try to discover the underlying one. 

We start by adding dimension 6 operators to the SM Lagrangian (dim 5 corresponds to the Weinberg operator, that does not appear in Higgs processes and dim$>6$ is out of reach for the current experimental thresholds) 

\begin{equation}
\lag_{eff} = \lag_{SM}^{(4)} + \frac{1}{\Lambda^2} \sum_k \alpha_k \op_k^{(6)} 
\end{equation}

It was shown in \cite{warsaw} that the basis of gauge-invariant, dimension 6, \textit{independent}\footnote{not related through the equations of motion. But, be careful: when looking at Green functions instead of S-Matrix elements, those equivalent operators may lead to different expressions} operators contains 59 terms (2499 if we introduce flavours). It is very important not to take an overcomplete set, since it would not satisfy the Ward-Slavnov-Taylor identities \cite{WST1, WST2, WST3}, on the contrary, it is possible to chose an undercomplete basis as long as one sticks to the relevant WST identities for it. In particular, we chose a 26-operator subset of the so-called Warsaw-basis \cite{warsaw}, containing only the relevant operators for the one loop renormalization of the Higgs sector and the Higss two-body decays. We do not consider for instance CP-even operators\footnote{Same as in the original kappa framework: a priori we only consider deviations that respect the tensor structures of the couplings. But note that in this case we do not lose generality by doing this: the remaining operators and their Wilson coefficients can be added any time later. See for instance \cite{manohar2}}. The list can be found in table \ref{table:26ops}.

\begin{table}  
\begin{center}
\begin{tabular}{l || l} 
\hline 
   $ \op_1 = g^3 \op_{\phi} = g^3 (\Phi^\dagger \Phi)^3$ & 
   $ \op_2 = g^2 \op_{\phi \Box} = g^2 ( \Phi^\dagger \Phi) \Box (\Phi^\dagger \Phi) $  \\ 
	$ \op_3 = g^2 \op_{\phi D} = g^2 ( \Phi^\dagger D_\mu \Phi) \left[ (D_\mu \Phi)^\dagger \Phi \right]$ & 
	$ \op_4 = g^2 \op_{\ell \phi} = g^2 (\Phi^\dagger \Phi ) \overline{\rm{L}}_{\rm{L}}  \Phi^c \ell_R$  \\
   $ \op_5 = g^2 \op_{u \phi} = g^2 (\Phi^\dagger \Phi)\overline{\rm{q}}_{\rm{L}}  \Phi \, \, \rm{u}_R $ &
    $ \op_6 = g^2 \op_{d \phi} = g^2 (\Phi^\dagger \Phi) \overline{\rm{q}}_{\rm{L}}  \Phi^c \, \, \rm{d}_R $  \\
   $ \op_7 = g^2 \op_{\phi \ell}^{(1)} = g^2 \Phi^\dagger D_\mu^{(\leftrightarrow)}  \Phi \overline{\rm{L}}_{\rm{L}} \gamma^\mu \rm{L}_{\rm{L}}$ & 
   $ \op_8 = g^2 \op_{\phi q}^{(1)} = g^2 \Phi^\dagger D_\mu^{(\leftrightarrow)}  \Phi \overline{\rm{q}}_{\rm{L}} \gamma^\mu \rm{q}_{\rm{L}} $  \\
   $ \op_9 = g^2 \op_{\phi \ell} = g^2 \Phi^\dagger D_\mu^{(\leftrightarrow)}  \Phi \overline{\ell}_{\rm{R}} \gamma^\mu \ell_{\rm{R}}$ &
    $ \op_{10} = g^2 \op_{\phi u} = g^2 \Phi^\dagger D_\mu^{(\leftrightarrow)}  \Phi \overline{\rm{u}}_{\rm{R}} \gamma^\mu \rm{u}_{\rm{R}} $  \\
   $ \op_{11} = g^2 \op_{\phi d} = g^2 \Phi^\dagger D_\mu^{(\leftrightarrow)}  \Phi \overline{\rm{d}}_{\rm{R}} \gamma^\mu \rm{d}_{\rm{R}}$ &
   $ \op_{12} = g^2 \op_{\phi u d} = i g^2 (\Phi^\dagger D_\mu \Phi ) \overline{\rm{u}}_{\rm{R}} \gamma^\mu \rm{d}_{\rm{R}}$  \\
	$ \op_{13} = g^2 \op_{\phi \ell}^{(3)} = g^2 \Phi^\dagger \tau^a D_\mu^{(\leftrightarrow)} \Phi \overline{\rm{L}}_{\rm{L}} \tau^a \gamma^\mu \rm{L}_{\rm{L}}$ & 
	$ \op_{14} = g^2 \op_{\phi q}^{(3)} = g^2 \Phi^\dagger \tau^a D_\mu^{(\leftrightarrow)} \Phi \overline{\rm{q}}_{\rm{L}} \tau_a \gamma^\mu \rm{q_L} $  \\
	$ \op_{15} = g \op_{\phi G} = g (\Phi^\dagger \Phi ) \rm{G}^{a \mu \nu} \rm{G}^{a}_{ \mu \nu}$ & 
	$ \op_{16} = g \op_{\phi W} = g (\Phi^\dagger \Phi ) \rm{F}^{a \mu \nu} \rm{F}^{a}_{ \mu \nu} $  \\
	$ \op_{17} = g \op_{\phi B} =  g (\Phi^\dagger \Phi ) \rm{F}^{0 \mu \nu} \rm{F}^{0}_{ \mu \nu}$ & 
	$ \op_{18} = g \op_{\phi W B} = g \Phi^\dagger \tau^{\rm{a}}  \Phi \rm{F}_{a}^{\mu \nu}  \rm{F}^{0}_{ \mu \nu} $  \\
	$ \op_{19} = g \op_{\ell W} = g \overline{\rm{L}}_{\rm{L}} \sigma^{\mu \nu} \ell_{\rm{R}}  \tau_a \Phi^c F^a_{\mu \nu}$ & 
	$ \op_{20} = g \op_{u W} = g \overline{\rm{q}}_{\rm{L}} \sigma^{\mu \nu} \rm{u}_{\rm{R}} \tau_a \Phi F^a_{\mu \nu}$  \\
	$ \op_{21} = g \op_{d W} = g \overline{\rm{q}}_{\rm{L}} \sigma^{\mu\nu} \rm{d}_{\rm{R}} \tau_a \Phi^c F^a_{\mu \nu} $ & 
	$ \op_{22} = g \op_{\ell B} = g \overline{\rm{L}}_{\rm{L}} \sigma^{\mu \nu}  \ell_{\rm{R}} \Phi^c F^0_{\mu \nu}$  \\
	$ \op_{23} = g \op_{\rm{u B}} = g \overline{\rm{q}}_{\rm{L}} \sigma^{\mu \nu} \rm{u_R} \Phi F^0_{\mu \nu} $ & 
	$ \op_{24} = g \op_{\rm{d B}} = g \overline{\rm{q}}_{\rm{L}} \sigma^{\mu \nu} \rm{d_R} \Phi^c F^0_{\mu \nu}$  \\
	$ \op_{25} = g \op_{\rm{u G}} = g \overline{\rm{q}}_{\rm{L}} \sigma^{\mu \nu} \rm{u_R} \lambda_a \Phi G^a_{\mu \nu} $ & 
	$ \op_{26} = g \op_{\rm{d G}} = g \overline{\rm{q}}_{\rm{L}}  \sigma^{\mu \nu}  \rm{d_R} \lambda_a \Phi^c G^a_{\mu \nu}$  \\
    \hline
\end{tabular}
\end{center}
\caption{Subset of dim = 6 operators relevant for our applications (Table 1 of \cite{NLOHEFT}). Observe that for every non-hermitian operator, its hermitian conjugate (times the complex conjugate Wilson coefficient) has to be included in order to make the Lagrangian hermitian.\protect\footnotemark}
\label{table:26ops}
\end{table}
The idea is to renormalize this Lagrangian and present an expression for $\lag$ that contains a pure SM part plus some deviations (dim. 6 operators and counterterms to make the theory UV finite at one loop). 
In order to separate ``new'' operators from the ``old'', SM, ones, we have to redefine the fields to recover canonical normalization: dimension 6 operators modify the quadratic part of the Lagrangian, for instance the Higgs \textit{vev} gets shifted: $ \nu^2 = \nu_{SM}^2 \left( 1 + \Delta \nu^2 \right)$, with $\Delta \nu^2 \propto \frac{1}{\Lambda^2}$.

\newpage
\footnotetext{The convention for quarks and leptons is: $ \rm{q_L} = \left( \begin{array}{c}
u \\
d \\
\end{array} \right)_L , \quad  \rm{L_L} = \left( \begin{array}{c}
\nu_{\ell} \\
\ell \\
\end{array} \right)_L $ with $u = \{ u, c, t \}$, $d = \{d, s, b \}$ , \newline $\ell = \{e, \mu, \tau \}$ and $\rm{f_{L,R}} = \frac{1}{2} (1 \pm \gamma^ 5) \rm{f}$.}


%

\subsection{Analytical structure of the Amplitude}

After having defined our $\lag_{\rm{dim=6}}$, we can  write down any NLO amplitude as a sum of pure SM contributions (one-loop), tree contact terms with one dimension 6 operator insertion, and one-loop diagrams with one dimension 6 operator insertion:

\begin{equation}
\amp = \sum_{n=N}^\infty g^n \amp_n^{(4)}  +  \sum_{n=N_6}^\infty \sum_{l = 0}^n \sum_{k = 1}^\infty g^n g_{4+2k}^{k + l} \amp_{ n l k }^{(4 + 2k)} .
\end{equation}

In particular, the dim 6 case (k=1) reads:
\begin{equation}
\amp = \sum_{n=N}^\infty g^n \amp_n^{(4)}  + g_6  \sum_{n=N_6}^\infty \sum_{l = 0}^n  g^n g_{6}^{l} \amp_{ n l}^{(6)} , \qquad  \rm{with:} \quad g_6 = \frac{1}{\sqrt{2}G_F \Lambda^2}  .
\end{equation}

N is the leading order of the process we are considering (N=1 for $H \rightarrow VV$, N=3 for $H \rightarrow \gamma \gamma $, etc.) and N$_6 = 1$  for tree-initiated processes and N$_6 = N-2$ for loop-initiated ones. 
Note that we do not include pure dimension 6 terms, only pure dimension 4 terms and the interference between dim. 4 and dim. 6.  Nevertheless the purely dimension 6 terms (squared) are very interesting: they can be used to estimate the theoretical uncertainty for our NLO calculation.

\section{Overview of the calculation}

\epigraph{``Renormalization is easy, you learned it when you were a kid''}{}

The renormalization procedure is the same as you learned for the standard model: add counterterms for fields and parameters, construct self-energies, Dyson resum them and make the propagators to be  UV finite.

\begin{equation}
\Phi = Z_\Phi , \qquad p = Z_p \, p_{ren}, \qquad Z_i = 1 + \frac{g^2}{16 \pi^2} \left( \rm{dZ}_i^{(4)} + g_6 \rm{dZ}_i^{(6)} \right)  
\end{equation}

Where $\Phi$ are the fields and $p$ parameters. Further, look at 3-point functions, and remove remanent UV divergences. In this particular case we found that Wilson coefficients need to be mixed in order to cancel such divergences:
 
 \begin{equation}
 W_i = \sum_j Z_{ij}^W W_j^{ren}
 \end{equation}

The mixing-coefficients can be found in Appendix G of \cite{NLOHEFT}. Renormalized Wilson coefficients are scale dependent, and the logarithm of the scale can be resummed in terms of the anomalous dimension matrix \cite{manohar1,trott2015, manohar2}.

\subsection*{Note: Range of applicability}

Field theories, up to some exceptions, have an energy scale. And it is a common mistake to think that this scale ($\Lambda, \mu_R , \mu_{\rm{QCD}}$) can take any value. For our Higgs EFT (NLO, dim 6) the range where we can be confident of our predictions is: $3 \, \, \rm{ TeV} < \Lambda <5 \,\, \rm{ TeV}$. Here we will try to make these thresholds clearer: 
\begin{itemize}
 \item $\Lambda > 3$ TeV :
Higher dimensional operators can be classified as PTG (potentially tree generated) or LG (loop generated), depending on their origin in the UV theory \cite{PTG}. Loop generated operators are suppressed by a factor of $1/16 \pi^2$ with respect to tree generated ones. Moreover, the Wilson coefficients for dim=8 tree-generated operators are of order $\approx \nu^2 /\Lambda^2 $. This means that for values of $\Lambda \leq 3$ TeV, we must either neglect loop-generated operators or include dimension 8 terms. 
 
\item $\Lambda < 5$ TeV: Look at the order of the new coupling constant, $g_6 = \frac{1}{\sqrt{2}G_F \Lambda^2} $. For values of $\Lambda$ around 5 TeV, the dimension 6 contributions are of the same size of the loop contributions: $g_6 \approx \frac{g^2}{4 \pi}$. If we choose the EFT scale to be bigger, we should also include higher-loop corrections to stay consistent.  
\end{itemize}
 
\section{Some results for Higgs 2-body decays}

Let us sketch here the $H \rightarrow \gamma \gamma$ processes commented in the beginning. The amplitude for the process $H(P) \rightarrow A_{\mu} (p_1) A_{\nu} (p_2)$ is:

\begin{equation}
A_{HAA}^{\mu \nu} = \mathcal{T}_{HAA} \frac{p_2^\mu p_1^\nu   - p_1 \cdot p_2 \delta^{\mu \nu}}{M_H^2}
\end{equation}

And $\mathcal{T}_{HAA}$ was found to be \cite{NLOHEFT}:

\begin{equation}
\mathcal{T}_{HAA} = i \frac{g^3}{16 \pi^2} \left( \mathcal{T}_{HAA}^{(4)} + g_6 
\underbrace{\mathcal{T}_{HAA}^{(6),b}}_{\rm{UV divergent}} \right) + i g g_6  \underbrace{ \mathcal{T}_{HAA}^{(6),a}}_{\rm{UV finite}},
\end{equation}

where we can recognize the purely standard-model part, $\mathcal{T}_{HAA}^{(4)}$, containing the bosonic and $t,b$ loops, and the new contributions coming from the dimension 6 insertions: $\mathcal{T}_{HAA}^{(6)}$.
As mentioned before, the counterterms are not enough to remove the divergent part. We need to mix the Wilson coefficients in order to remove remaining UV divergences. Finally the renormalized amplitude can be written as:

\begin{equation}
\mathcal{T}_{HAA}^{ren} = i \frac{g_{ren}^ 3}{16 \pi^2} \left( \mathcal{T}_{HAA}^{(4)} + g_6 \mathcal{T}_{HAA;fin}^{(6),b} + \mathcal{T}_{HAA}^{(6),R} \ln\frac{\mu_R^2}{M^2} \right) + i g_{ren} g_6 \mathcal{T}_{HAA}^{(6),a}
\end{equation}

The scale dependence has been left explicit, such that one can compare this result with the one in \cite{trott2015} and find out they agree, or use the renormalization group equations to resum these logarithms. The anomalous dimension matrix for the complete basis\footnote{2499 operators} of SM dimension 6 operators was calculated in \cite{manohar1}.

Note that up to now we did as few approximations as possible. In particular, we did not apply the zero-width approximation, and we did not neglect loop-generated operators. In this particular case we could now neglect such LG operators to find that all non-factorizable terms vanish.

\section{Conclusions}

We discussed a method to study SM deviations from a SM point of view, at next-to-leading order. From a phenomenological point of view, the current challenge is to see how can we take the best out of the data from LHC RUN-II, given the change of paradigm with respect to RUN-I: after the Higgs has been observed we are not searching for a specific resonance anymore.

We strongly believe that the combination of EFT + POs will help us shed some light on the possible BSM scenarios. Of course specific BSM models are important and interesting to study, but we also  need a tool to explore them  within the current experimental thresholds. 


Having said that, the next challenge is to define the experimental strategy to follow. Some suggestions regarding a Bayesian analysis of Higgs couplings have been presented in \cite{Blas1}. In \cite{trottLEP} a number of pseudo-observables are fitted against pre-LHC data, as a first test for SM-EFT.  A discussion on how to approach this question, namely how to use electroweak precision data and other well-known measurements (electric dipole moment, masses, \dots) in the context of NLO-EFT is addressed in \cite{GPAndre}.

Moreover, the software \textit{Rosetta} \cite{rosetta} has been recently launched. With it, one can translate different effective operators into the main dimension 6 bases. Then, by means of \textit{Feynrules} \cite{feynrules},  one can implement any EFT into various analysis tools, such as Monte Carlo generators, that can be used to fit LHC data.

\bibliography{NewLibrary}{}
\bibliographystyle{JHEP}

\end{document}